\documentclass[fleqn,usenatbib]{mnras}

\usepackage{newtxtext, newtxmath}

\usepackage[T1]{fontenc}

\DeclareRobustCommand{\VAN}[3]{#2}
\let\VANthebibliography\thebibliography
\def\thebibliography{\DeclareRobustCommand{\VAN}[3]{##3}\VANthebibliography}

\usepackage{graphicx}
\usepackage{amsmath}
\usepackage{xspace}
\usepackage{pgfplots}
\usepackage{cleveref}
\usepackage{gensymb} %
\usepackage{multirow}
\usepackage{caption}
\usepackage{bm}
\usepackage{threeparttable}
\pgfplotsset{compat=1.15}

\pgfplotsset{
  discard if/.style 2 args={
    x filter/.code={
      \edef\tempa{\thisrowno{#1}}
      \edef\tempb{#2}
      \ifx\tempa\tempb
      
      \fi
    }
  },
  discard if not/.style 2 args={
    x filter/.code={
      \edef\tempa{\thisrowno{#1}}
      \edef\tempb{#2}
      \ifx\tempa\tempb
      \else
      
      \fi
    }
  }
}

\usepgfplotslibrary{dateplot}
\usetikzlibrary{external}
\newcommand{\insitu}{\textit{in situ}\xspace}
\newcommand{\Insitu}{\textit{In situ}\xspace}
\newcommand{\vecc}[1]{\mathbf{#1}}

\newcommand{\ngroup}{n_{\mathrm{gr}}}
\newcommand{\Ne}{N_{\mathrm{e}}}
\newcommand{\Np}{N_{\mathrm{p}}}
\newcommand{\Na}{N_{\mathrm{a}}}
\newcommand{\Ncrit}{N_{\mathrm{crit}}}

\newcommand{\dd}{\mathrm{d}}

\title[Solar Wind Density Model for Spacecraft Ranging]{Improving the Solar Wind Density Model Used in Processing of Spacecraft Ranging Observations}

\author[Dan Aksim, Dmitry Pavlov]{
Dan Aksim,$^{1}$\thanks{E-mail: danaksim@iaaras.ru}
Dmitry Pavlov,$^{2}$
\\
$^{1}$Institute of Applied Astronomy, St. Petersburg, Russia\\
$^{2}$St. Petersburg Electrotechnical University, Russia\\
}

\date{Accepted XXX. Received YYY; in original form ZZZ}

\pubyear{2022}

\usepackage{xcolor, soul}
\sethlcolor{green}

\begin{document}
\label{firstpage}
\pagerange{\pageref{firstpage}--\pageref{lastpage}}
\maketitle

\begin{abstract}
    Solar wind plasma as a cause of radio signal delay has been playing an
    important role in solar and planetary science. Early experiments
    studying the distribution of electrons near the Sun from spacecraft
    ranging measurements were designed so that the radio signal was
    passing close to the Sun. At present, processing of spacecraft
    tracking observations serves a different goal: precise (at meter
    level) determination of orbits of planets, most importantly Mars.
    Solar wind adds a time-varying delay to those observations, which
    is, in this case, unwanted and must be subtracted prior to
    putting the data into planetary solution. Present planetary
    ephemeris calculate the delay assuming symmetric stationary power-law model
    of solar wind density. The present work, based on a custom variant
    of the EPM lunar-planetary ephemeris, raises the question of accuracy
    and correctness of that assumption and examines alternative
    models based on \textit{in situ} solar wind density data provided by OMNI and
    on the ENLIL numerical model of solar wind.
\end{abstract}


\section{Introduction}
\label{sec:intro}

Two-frequency observations of the Viking spacecraft in 1976 (lander
and orbiter) were once used for analysis of radio signal propagation delays
due to solar wind plasma
\citep{Callahan1977,Muhleman1981}.
Also, there were plans for dual-frequency ranging for the Mars Express
spacecraft~\citep{Patzold2004,MEX-Corona}, but no information is known about
the outcome.
Apart from those, the spacecraft tracking measurements are
made on single frequency, thus determining the delay in solar wind
directly from the measurements themselves is not possible. While, in
principle, other radio observations could help, such as
VLBI \citep{Aksim2019} or observations of scintillations or radio
sources \citep{Shishov2016}, in practice there are too few of them, and
even fewer, if any, on the dates of spacecraft tracking measurements.
Three options remain for accounting for delay in solar wind when determining
planetary orbits for ephemeris:
\begin{enumerate}
\item Using a simplified model of solar wind electron density distribution,
  and then determining its single free parameter (a reference electron density) from
  the spacecraft tracking observations. 
\item Using \insitu electron density measurements done by
  spacecraft such as the Advanced Composition Explorer (ACE) and
  \textit{Wind}, to account for large-scale
  temporal variations of the said electron density in the
  interplanetary space, while keeping a simplified symmetric model for
  electron density distribution at any moment.
\item Using a more complicated dynamical model of solar wind, which
  accounts for corotating interaction regions existing in solar wind.
  Such models of the solar wind are often built on top of models
  of solar corona, whose parameters are fit to magnetograms, either
  daily or mean (averaged over one solar cycle). This approach
  does not necessarily preclude using the \insitu data.
\end{enumerate}

The symmetric models most often include the term of electron density
proportional to $1/r^2$, where $r$ is the distance to the center of the
Sun. This term assumes constant radial velocity of electrons%
\footnote{
  The connection between density distribution $\Ne(r)$ and radial velocity
  $v_r(r)$ is expressed by the equation $r^2 v_r(r) \Ne(r) = \mathrm{const}$,
  which follows from the continuity equation \citep{Aksim2019}.
}.
The factor of this term (reference electron density) is a free
parameter of the model that is determined from spacecraft ranging
observations.

While all modern planetary ephemeris (DE, EPM, INPOP) use the $1/r^2$
density, their approaches differ in details. DE200
\citep{Standish1990} used the following model for solar wind
correction of Mariner ranges:
$$ \Ne = \frac{A}{r^6} + \frac{B}{r^2},$$
\noindent where $A$ is a fitted parameter and $B$ depends on
solar latitude $\beta$:
$$B = \frac{ab}{a^2 \sin^2\beta + b^2 \cos^2\beta}$$
\noindent ($a$ and $b$ are fitted parameters as well).
DE430 \citep{DE430TR} presumably uses a similar model, where the
parameters are determined per-planet or per-spacecraft.

In EPM2004 \citep{Pitjeva2004} and subsequent releases of EPM, the following
model was used:
\begin{equation}\label{eq:bbdot}
  \Ne = \frac{A}{r^6} + \frac{B + \dot{B} t}{r^2},
  \end{equation}
\noindent where $A$ was fixed to the value determined in DE200, and
$B$ with its linear drift $\dot{B}$ were determined from observations,
per-planet, per-year.

INPOP presumably has two models that are used interchangeably
\citep{Verma2013}: (i) a two-parameter model with one term proportional
to $1/r^2$ and the other proportional to $1/r^4$; (ii) a two-parameter
model with the term proportional to $1/r^\epsilon$, where $\epsilon$
is the second determined parameter. The parameters are fit per each
solar conjunction of each planet of interest, and also separately for
spacecraft being in the fast-wind zone and in the slow-wind zone.

Outside planetary ephemeris, other solar wind electron density
models exist; see e.g. \citet{Leblanc98} where $1/r^2$, $1/r^4$, and
$1/r^6$ terms are combined.  However, the $1/r^4$ and higher order
terms have negligible effect on spacecraft tracking signals that pass
farther than 15 solar radii ($15 R_{\odot}$) from the center of the Sun.
The signals that pass closer than that cannot be used for precise orbit
determination anyway because of the temporal and spatial instabilities
of the delay in solar wind in such proximity.

There exist other models with fractional powers of $r$ close to 2, see
e.g. \citet{Bougeret1984} with $r^{-2.10}$, \citet{Mann1999} with
$r^{-2.16}$, or \citet{Verma2013} with values of $\epsilon$ ranging
from 2.10 to 2.50. Such models in general do not possess physical
meaning and, as we will now show, do not offer prospects of more
accurate representation of spacecraft signal delay.

A major problem in determination of solar wind parameters from
spacecraft ranges (aside from the temporal, latitudinal, and longitudinal
variations) is that not only the factor of the $1/r^2$ (or
$1/r^\epsilon$) term must be determined from observations, but the
whole set of parameters that form the ephemeris solution. Those
include parameters of orbits of the planets, mass of the Sun, masses
of dozens of asteroids, and, most importantly, signal delays (biases)
that come from spacecraft transponder going out of calibration and
from systematic errors of signal registration in radio observatories
\citep{Kuchynka2012}.

In fact, just having one factor and one offset to fit for the 
delay of spacecraft ranges in solar wind already makes the power of $r$
largely irrelevant. \Cref{fig:1r2_vs_1r25} in \Cref{sec:difficulties}
provides an example where $1/r^{2.5}$ model is fitted to data generated
using the $1/r^2$ model well enough for practical purposes.

We emphasise that there is no power law that can be preferable in a
symmetric model of solar wind electron density --- either from the
position of physical principles or from the position of processing of
signal delays.

\subsection{Medium-term density variations}\label{sec:meduim-term}

Medium-term variations in the solar wind are ones that have duration similar
to the period of solar rotation (Sun's synodic rotation period is 27.2753 days).
One obvious problem with the approach of using the $1/r^2$ model
is the assumption that solar wind is radially symmetric.
The asymmetry of the solar wind can be clearly seen, for instance,
in the LASCO coronagraph images%
\footnote{\url{http://www.swpc.noaa.gov/products/lasco-coronagraph}}.
The fact that spacecraft ranging
observations are sensitive enough to be affected by this asymmetry
was reported by \citet{Kuchynka2012}, who analyzed 10-day rolling averages
of the observation residuals and noticed the presence of periodic variations
with a period of 27 days. (See \Cref{sec:residuals} for brief introduction
to radio ranging residuals).

The way to account for these periodic variations is to adopt, instead of the
$1/r^2$ model, a three-dimensional numerical model of the solar wind that
would simulate (to some level of agreement) the corotating structures of the solar wind
based on observations of the Sun.

In this work, we examine the results obtained with the WSA-ENLIL model,
which consists of two parts:
the ENLIL solar wind model \citep{Odstrcil2003} and
the WSA \citep[Wang--Sheeley--Arge, ][]{Wang1990,Arge2000,Sheeley2017} model
of the solar corona.

ENLIL is a magnetohydrodynamic (MHD) solar wind model that calculates
3D distributions of solar wind's density, velocity, and magnetic field
by solving a system of differential MHD equations using a coronal solution
given by WSA as a boundary condition at 0.1~AU.
WSA is a semi-empirical solar corona model that uses line-of-sight
magnetic field measurements of the solar photosphere assembled into synoptic
magnetograms and simulates the solar wind plasma outflow up to 0.1~AU.

\begin{figure}
  \centering
  \includegraphics[height=5cm]{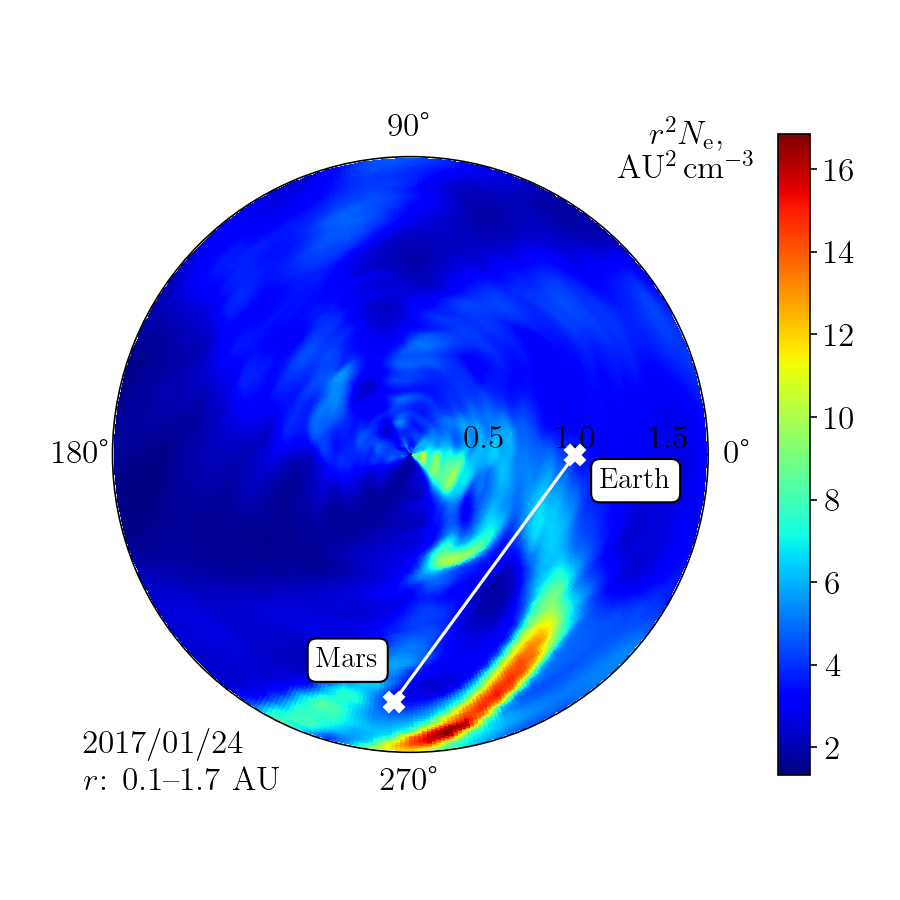}

  \includegraphics[height=5cm]{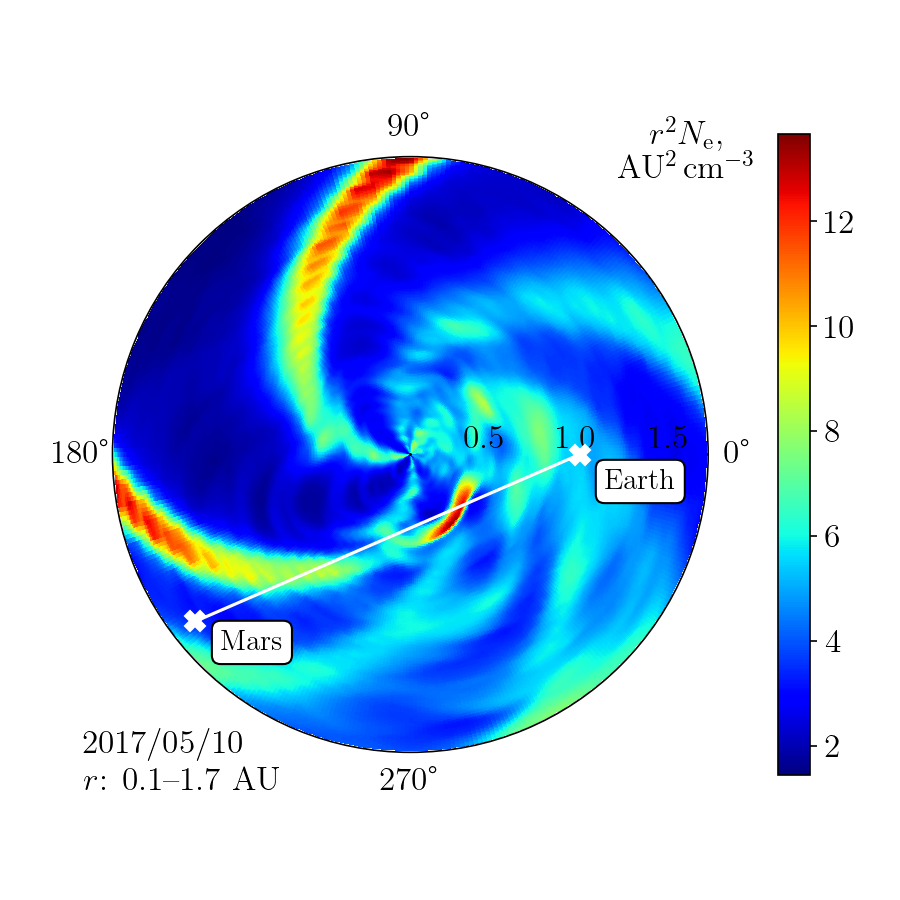}
  \caption{
    Normalized solar wind electron number density calculated by
    WSA-ENLIL with daily-updated GONGz magnetograms.
    The data are plotted in the helio-equatorial plane with latitude of 0\degree\ in
    the HEEQ coordinate system \citep[see][]{Thompson2006}.
  }
  \label{fig:enlil_sample_map}
\end{figure}

The primary purpose of WSA-ENLIL is prediction of solar wind
structures reaching Earth (and, with the help of the CONE model,
prediction of geomagnetic storms caused by coronal mass
ejections). However, there is no reason the model cannot be used for
other purposes, such as processing of spacecraft ranging observations.

\Cref{fig:enlil_sample_map} shows WSA-ENLIL density distributions
in the solar equatorial plane for January 24 and May 10, 2017.
The spiral shapes in the map are corotating solar wind structures.

\subsection{Long-term density variations}

Along with previously mentioned 27-day periodic variations in solar wind's
density, there are also long-term variations that happen on a larger
scale, i.e. on a scale of years.
These long-term variations can be seen by analysing \insitu density
measurements applying a rolling average filter with a window size
large enough to smooth out any medium-term data variations.

\begin{figure}
  \centering
  \includegraphics{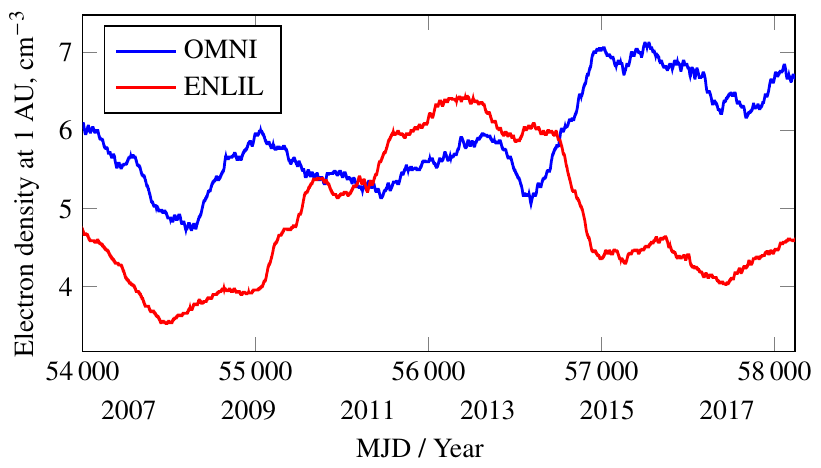}

  \caption{
    Solar wind electron density at 1 AU (365-day rolling average),
    years 2006--2017.
  }
  \label{fig:density_1au}
\end{figure}

\Cref{fig:density_1au} shows 365-day rolling average of the solar wind electron
density at 1~AU calculated twice: (i) from the OMNI dataset
\citep{OMNI2014,WindACE2005} using values of proton density and alpha--proton
ratio and assuming charge neutrality (see \Cref{sec:insitu-data}),
and (ii) from the ENLIL simulations (see \Cref{sec:enlil_data}).
The plot reveals that the average electron density in the long term can deviate
up to 20\% from its mean value, so the assumption of some ``nominal'' electron
density for all time would be unjustified in the processing of spacecraft radio ranges.

Our suggestion is that these long-term variations must be taken into account
regardless of the particular solar wind model used.
Specifically, if the power-law model like $N_0/r^2$ is used,
one constant $N_0$ multiplier would result in constant density at 1~AU, which
would not be in agreement with \insitu observations.
Therefore, $N_0$ must be made proportional to \insitu density smoothed in
a way that would eliminate any medium-term disturbances and variations,
such as 27-day periodic ones resulting from the Sun's rotation.

In the case of ENLIL, we have to take into account the inner workings of the
model~\citep{ENLILCode}.
First, we must note that ENLIL assumes that the electron and proton densities are equal.
Also, it is important that to calculate the density
at the inner boundary, ENLIL requires that at $r = 0.1$~AU the condition
of constant momentum flux holds:
\begin{equation}
  \Ne(\vartheta, \varphi)\, v_r^2(\vartheta, \varphi) =
      d_\mathrm{fast}^{\phantom{1}} \cdot v_\mathrm{fast}^2 = \mathrm{const},
\end{equation}
in which $v_r(\vartheta, \varphi)$ is the radial velocity distribution produced
by the WSA solar corona model, $\Ne(\vartheta, \varphi)$ is the density
distribution further used as the boundary condition, and
$d_\mathrm{fast}$ and $v_\mathrm{fast}$ are free model parameters named
\texttt{dfast} and \texttt{vfast}.
Presumably, parameters $d_\mathrm{fast}$ and $v_\mathrm{fast}$ should be
adjusted so that the solution matches observational data.
If great precision is not a requirement, this adjustment may be performed
post factum, i.e. by directly scaling the solution $\Ne(r, \vartheta, \varphi)$.

The WSA-ENLIL average density curve shown in \Cref{fig:density_1au}
was calculated with fixed values of $d_\mathrm{fast}$ and $v_\mathrm{fast}$,
i.e. with one constant value of momentum flux.
The figure shows that when momentum flux is fixed, WSA-ENLIL's average electron
density at 1 AU (i) is not constant and (ii) does not match the OMNI numbers.

Quite obviously, with average density being non-constant, simply setting
$d_\mathrm{fast}$ proportional to electron density derived from OMNI would do very little
to bring the two curves in agreement.
Instead, knowing that variations in density are defined by variations
in WSA velocity, we presume that the first step should be to use values of
$v_\mathrm{fast}$ based on WSA velocity distribution $v_r(\vartheta, \varphi)$,
possibly choosing the distribution's average as the parameter's value.
Doing so would produce solutions with near-constant average density at 1~AU,
and, at that stage, $d_\mathrm{fast}$ may be set to values proportional to OMNI
density. Such a setup should, in theory, produce solutions that match
observational \insitu data.

However, since ENLIL simulations require time to complete and
solutions may be scaled post factum, the simplest (and the one used in this work)
way to bring ENLIL data in agreement with OMNI is to remove
the density trend shown in \Cref{fig:density_1au} directly from the solution
$\Ne(t, r, \vartheta, \varphi)$ and to apply to the data the same procedure
as we do to $N_0$ parameter of the $N_0/r^2$ model.

For a detailed description of the \insitu\ data used, see \Cref{sec:insitu-data}.

\section{Radio signal propagation in solar wind}

Radio signal propagation velocity in plasma is defined by
the group refractive index \citep{Aksim2019}
\begin{equation}
  \label{eq:n_group}
  \ngroup(\vecc r, \omega) \approx 1 + \frac{\Ne(\vecc r)}{2 \, \Ncrit(\omega)},
\end{equation}
where $\omega$ is the signal's angular frequency,
$\Ne(\vecc r)$ is the plasma's electron number density,
and $\Ncrit(\omega) = m_\mathrm{e} \, \varepsilon_0 \, \omega^2/e^2$ is
critical plasma density ($e$ is the charge of electron, and $\varepsilon_0$
is the vacuum permittivity).

To calculate the time%
\footnote{Conforming to the tradition of measuring time delays in units of length,
by ``time delay $\tau$'' we mean not \emph{time delay} specifically,
but rather \emph{optical path length}, which is measured in meters.}
it takes for a radio signal of frequency $\omega$ to travel
through the solar wind plasma from the point given by position vector $\vecc r_a$
to the point $\vecc r_b$, one has to integrate the plasma's group refractive
index $\ngroup(\vecc r, \omega)$ along the straight line connecting $\vecc r_a$
and $\vecc r_b$, that is, to evaluate the integral
\begin{equation}
  \label{eq:tau_integral}
  \begin{split}
  \tau &= \int_{\vecc r_a}^{\vecc r_b} \ngroup(s, \omega) \, \dd s = \\
       &=
         |\vecc r_b - \vecc r_a | +
         \frac{1}{2 \, \Ncrit(\omega)}\int_{\vecc r_a}^{\vecc r_b} \Ne(s) \, \dd s
         = \tau_\mathrm{dist} + \tau_\mathrm{cor}.
  \end{split}
\end{equation}

The term $\tau_\mathrm{dist}$ in the above expression corresponds
to the time delay
due to the distance between $\vecc r_a$ and $\vecc r_b$,
i.e. the time it would take for the signal to travel the same path in vacuum,
and $\tau_\mathrm{cor}$ is the additional dispersive time delay due to the lower
propagation velocity in solar wind.

In the case of a power-law electron density model $\Ne(r) \propto r^\alpha$
the integral in \Cref{eq:tau_integral} may be expressed in closed form
\citep[see][]{Aksim2019}.
In the general case, specifically if the electron density is defined by a grid
of numerical values, which is the case with ENLIL and other numerical models,
the integral has to be evaluated numerically.
To calculate the integral numerically, we used the Simpson's rule
\citep[sec. 5.1]{Kahaner1989} with step size of 0.00625~AU,
equal to the radial grid spacing of the used simulation data
(see \Cref{sec:enlil_data}), which allows to exclude possible error
originating from the grid.

\section{Data}
Despite this work being focused on Mars spacecraft observations, each
of the considered solar wind models was tested as part of its
respective planetary solution. Each solution was fit not only to Mars
spacecraft observations, but to the full set of observations usually
required to build ephemeris, including not only spacecraft ranging, but
also optical observations and other kinds of
observations. For details, we refer the reader to \citep{Pitjeva2013,PitjevaPitjev2014}.

\subsection{Spacecraft ranges}\label{sec:spacecraft-ranges}

The Mars spacecraft ranges that were crucial to this work were taken
from the webpage of Solar System Dynamics (SSD) group at NASA
JPL\footnote{\url{https://ssd.jpl.nasa.gov/planets/obs_data.html}}, which includes
Mars Global Surveyor (MGS, 1999--2006), 
Odyssey (2002--2013), and Mars Reconnaissance Orbiter (MRO, 2006--2013).
The extended set of ranges for MRO and Odyssey up to
the end of 2017 was kindly provided by William Folkner. The provided
ranges are normal points (NPs) made from raw observations. A normal point
represents the distance from an Earth radio observatory to the center of Mars,
corrected for the delay in ionosphere and troposphere.

Following the decision from \citep{Kuchynka2012}, pre-2005 Odyssey and
MRO ranges acquired by Deep Space Network stations 26 and 43 were
removed from analysis, as well as ranges acquired by station 54 before
October 2008.

MGS, Odyssey, and MRO ranges that pass closer than 30$R_{\odot}$
to the Sun suffer too much from solar wind instabilities and have
been removed from analysis. Ranges that pass farther than
30$R_{\odot}$, but closer than 60$R_{\odot}$ have been treated
separately (but equally with other ranges).

Mars Express (MEX) and Venus Express (VEX) ranges \citep{MorleyBudnik} that are
also sensitive to the delay in solar wind were downloaded from the Geoazur
website\footnote{\url{http://www.geoazur.fr/astrogeo/?href=observations/base}}.
They too are published as observatory--Mars distances corrected for
ionosphere and troposphere, but otherwise they are ``raw'' observations, so normal
points had to be made from them before processing. VEX ranges between 23.08.2010 and
31.10.2010 were excluded from analysis due to poor residuals, the reason
believed to be bad orbit determination due to trajectory control
maneuvers being more frequent than usual\footnote{This information
was obtained in personal communication from Elena Pitjeva, who
in turn obtained it from Trevor Morley.};
VEX ranges between 02.10.2011 and 05.12.2011 were excluded too,
assuming the same reason. Due to generally big noise in VEX residuals,
the decision was made to exclude VEX ranges that pass closer
than $90R_{\odot}$ to the Sun, so that they do not interfere
with less noisy observations in the comparison of the solar wind models.
The MEX ranges were treated similarly to Odyssey and MRO,
with the cutoff values of $30R_{\odot}$ and $60R_{\odot}$. 

MESSENGER \citep[2011--2014, ][]{Park2017} and Cassini (2004--2014)
spacecraft ranges (normal
points) were also taken from the aforementioned SSD webpage; however,
the published values have been already reduced for the delay in solar wind
(calculated by the JPL DE solar wind model), so they did not
participate directly in the analysis of solar wind models. However,
those observations are very important for Solar system ephemeris as a
whole; in particular, they reduce the uncertainty of the mass of the
Sun.

In this work, different models of solar wind plasma electron density are
compared in their application to building planetary ephemeris. One of the models,
WSA-ENLIL, depends (in its WSA part) on the GONG magnetograms
that did not exist before September 2006 (see
\Cref{sec:enlil_data}), so the comparison is done for 2006--2017.  The
spacecraft observations that touch this time span are summarized in
\Cref{tbl:data-spacecraft}. The reception (Rx) and transmission (Tx)
frequencies are listed so as to emphasise that the frequency
is important for calculation of the delay of radio signal in plasma,
see \Cref{eq:n_group}.

\begin{table*}
  \begin{threeparttable}
    \caption{Spacecraft observations used in the comparison of solar wind models.
             NPs = normal points, Rx = reception, Tx = transmission,
             $\sigma$ = \textit{a priori} error}
    \label{tbl:data-spacecraft}

    \begin{tabular}{lcrrrcc}
        \hline
        \textbf{Spacecraft} &
        \textbf{Time span} &
        \multicolumn{2}{c}{\textbf{\# of NPs}} &
        \textbf{$\sigma$, one-way} &
        \multicolumn{2}{c}{\textbf{Frequency}} \\
        & &
        \textbf{$> 60 R_{\odot}$} & \textbf{$30..60 R_{\odot}$} &
        &
        \textbf{Rx} & \textbf{Tx} \\
        \hline
        Odyssey & 2002--2017 & 7988 & 1784 & 0.5 m & 7155 MHz & 8407 MHz \\
        MRO     & 2006--2017 & 1924 & 561  & 0.5 m & 7183 MHz & 8439 MHz \\
        MEX     & 2005--2015 & 2888 & 718  & 1.5 m & 7100 MHz & 8400 MHz\\
        VEX     & 2006--2013 & 1294$^\dagger$\hspace{-0.42em} & ---  & 3 m & 7166 MHz & 8419 MHz \\ \hline
        \hline
        \multicolumn{7}{l}{
          $^\dagger$ Sun--signal distance for all VEX normal points
                     is greater than $90 R_\odot$.
        }
    \end{tabular}
  \end{threeparttable}
\end{table*}

\subsection{\Insitu solar wind measurements}
\label{sec:insitu-data}

Low-resolution (daily) version of the NASA/GSFC's OMNI
dataset\footnote{\url{http://spdf.gsfc.nasa.gov/pub/data/omni/low_res_omni}}
was used as the source of \insitu\ data in this work. The dataset is
combined\footnote{Methods used for the combination are outlined at
  \url{http://omniweb.gsfc.nasa.gov/html/omni_min_data.html}} from
\insitu data measured by multiple spacecraft. The ones that serve as sources
of the particle density
data for 2006--2017 are the ACE \citep{ACE1998}
and \textit{Wind} \citep{Wind2021}.

OMNI provides proton density $\Np$ and alpha--proton ratio $\Na/\Np$.
To calculate electron density $\Ne$,
we used the plasma quasi-neutrality condition $\Ne = \Np + \Na/2$.

\begin{figure}
  \centering
  \includegraphics{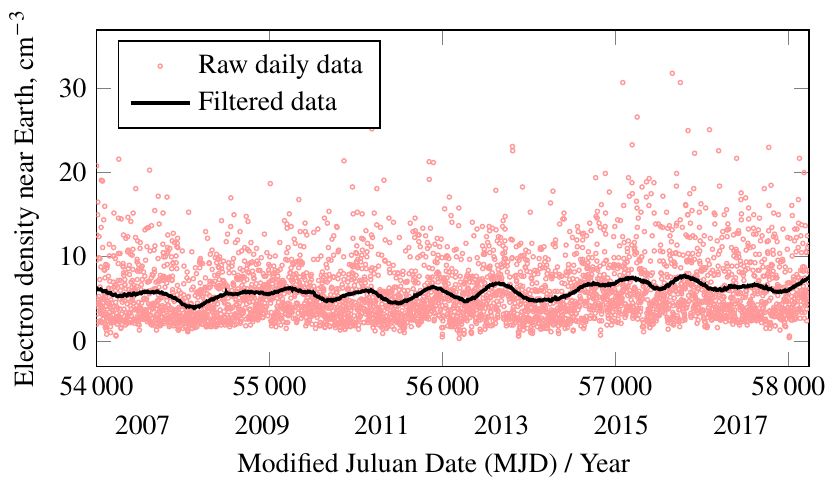}

  \caption{
    OMNI data of electron density near Earth, years 2006--2017. The filter used is Savitzky--Golay
    with a window of 511 days and the polynomial order 5.
  }
  \label{fig:omni-savgol}
\end{figure}

The daily density data provided by OMNI are volatile because of the
nature of the solar wind structures; see the ``raw'' electron density
data derived from OMNI on \Cref{fig:omni-savgol}. Since we are
interested in the (average) electron density along the signal path,
our analysis can not benefit from the information about the daily
variations at the end of the signal path (near Earth). On the other
hand, we assume that a sufficiently smoothed electron density curve
will represent long-term variations of electron density of the solar
wind in the whole inner Solar system, including the Earth--Mars line.
The Savitzky--Golay filter \citep{SavGol1964} was chosen as one of the
most widely used filters for time series. The parameters of the filter
were chosen in an ad hoc manual procedure: the window of 511 days and
the polynomial order 5. The filtered data are shown on \Cref{fig:omni-savgol}.

Alternatively to OMNI, \insitu\ electron density data are also provided
by the ESA/NASA Solar and Heliospheric Observatory \citep[SOHO,
][]{hovestadt1995celias}.
This data\footnote{\url{http://umtof.umd.edu/pm/crn}} is not part
of the OMNI combination.
The results obtained with the SOHO data (not shown in the text) turned out
to be somewhat worse than the ones obtained with OMNI.

\subsection{Numerical solar wind density maps}
\label{sec:enlil_data}

WSA-ENLIL data were produced by request via the ``Runs on Request'' service on
the Community Coordinated Modeling Center (CCMC)
website\footnote{\url{http://ccmc.gsfc.nasa.gov}}.
The whole dataset covers the dates between September 2006 and December 2017
and consists of 12 separate runs, each one covering one year.

The grid size in the WSA-ENLIL runs was set to
$256 \times 10 \times 90\ (r \times \vartheta \times \phi)$
with grid boundaries 0.1...1.7~AU, $-$20\degree...+20\degree\ latitude,
0\degree...360\degree\ longitude (in HEEQ coordinate system).
The grid spacing, therefore, is 0.00625~AU in radial direction and 4\degree\ 
in latitudinal and longitudinal directions.
The output frequency for the simulations was set to once every 12 hours.

The solar surface magnetic field data used for WSA coronal solutions
consisted of daily updated synoptic maps produced by the Global Oscillation
Network Group \citep[GONG, ][]{Hill2018GONG}.

It must be noted that there are two kinds of GONG magnetograms, called
GONGb and GONGz.  They were supposed to be identical with the only
difference that GONGz has non-solar magnetic field bias
removed. However, it was found in late 2015 that GONGz
magnetograms suffered from over-subtraction in the polar regions. The
correction of the over-subtraction gradually went into effect starting
from August 2016 and reaching all GONG stations in April
2017. Unfortunately, it is not possible to retroactively correct the
past GONGz data~\citep[sec. 5.4]{CCMCReport}. So the numerical
simulations in 2017 and later are expected to be more accurate when
GONGz is used; before 2017, the expected improvement of GONGz is not
reached and GONGb is more accurate.

\section{Analysis}
As said above, a full planetary ephemeris solution were obtained for
each of the considered solar wind models. All solutions were based on
EPM and ERA-8 software \citep{PavlovSkripnichenko}.

\subsection{Dynamical model}
\label{sec:dynmodel}
EPM has a single dynamical model of the Solar system that includes all
planets, the Moon, the Sun, asteroids (see
section~\ref{sec:asteroids}), discrete model of Kuiper belt
(160 uniformly distributed points of equal mass)
and 30 individual trans-Neptunian objects \citep{PitjevaKuiperBelt}.
Sixteen bodies
(the Sun, the planets, Pluto, Ceres, Pallas, Vesta, Iris, Bamberga)
obey Einstein--Infeld--Hoffmann equations of motion.  Other bodies,
for the sake of performance, interact with those 16 bodies with only
Newtonian forces, and do not interact with each other. It has been
checked that taking into account the full set of relativistic
interactions affects the Earth--Mars distance by a few centimeters,
which is negligible with the present observations.

Apart from point-mass interactions, the model includes additional
accelerations from solar oblateness and Lense--Thirring effect.
Earth also gets ``point mass--figure'' accelerations that come from the Sun,
Venus, Mars, Jupiter, and the Moon \citep{pavlov2016}.

The planetary part of the EPM model has over 600 parameters
that are determined simultaneously in the least-squares
fit (see \Cref{sec:residuals}).
Here are the parameters that are important for the analysis of
radio ranges from Earth to Mars orbiters:
\begin{itemize}
\item orbital elements for all planets at epoch,
\item solar oblateness factor,
\item masses of 279 asteroids,
\item distance scale parameter, historically known as correction to the value
      of AU,
\item single factor of electron density, which applies to delay in solar wind plasma,
\item biases to compensate for calibration errors or clock offsets on Earth
      or on spacecraft: 15 per-station biases for MGS/Odyssey, 14 per-station
      biases for MRO, single transponder delay for MGS, single transponder delay
      for Mars Express%
      \footnote{It is not possible, nor necessary, to determine a transponder
                delay for MRO, because it is absorbed by the per-station biases
                in the solution. For the same reason, of the MGS/Odyssey pair,
                only the MGS transponder delay is determined.}.
\end{itemize}

The distance scale parameter represents the correction to $GM_{\odot}$---the mass of
the Sun. It is numerically harder to determine the mass of the Sun
directly because of correlations with semi-major axes of planets.
Determination of $GM_{\odot}$ (and, in specific solutions, its time derivative)
allows to study the physical properties of the Sun and to test general
relativity \citep{Pitjeva2021}.

It is important to process different kinds of observations
simultaneously to get realistic values of parameters and error
estimates. For example, MESSENGER observations help greatly to
determine the solar oblateness factor, because Mercury is close to the
Sun. The solar oblateness factor affects orbits of Earth and Mars.  To
determine the distance scaling factor, one has to take ranging data
for MESSENGER, MEX, Mars orbiters, and Cassini. Earth's orbit, whose
precision is obviously important for processing of the Earth--Mars
ranges, depends on the Moon's orbit around Earth, which is determined
from lunar laser ranging observations. Optical observations are
important for determination of orbits of Uranus and Neptune,
who in turn affect Mars's orbit on long time spans.

\subsection{Perturbing asteroids}\label{sec:asteroids}
An important part of the model, especially in regard to the Mars
spacecraft observations, are the point-masses that represent the
asteroids in the Main Asteroid Belt.  In EPM2017, 301 largest
asteroids were present in the dynamical model; the remaining part of
the Main Asteroid Belt was modelled as a discrete rotating annulus
consisting of 180 uniformly distributed points of equal mass
\citep{Pitjeva2018,EPM2017}.  Masses of 16 individual asteroids were
fixed to the values determined from observations of spacecraft or
natural satellites orbiting them; 30 masses were determined as part of
ephemeris solution (i.e. by the perturbations that they inflict on
orbits of inner planets); other asteroids' masses were determined
as mean densities of three taxonomic classes (C, S, M).

In the updated ephemeris dynamical model used in this work, the number
of individual asteroids is 279. The source list of asteroids was
merged (with obvious removal of duplicates) from 343 asteroids of the
DE430 model \citep{DE430TR} and 287 asteroids from \citep[Table
  A.1]{Kuchynka2010}. The former is believed to be the list of most
perturbing asteroids, while the latter is believed to be the list of
most ``non-ring-like-acting'' asteroids, i.e. their cumulative effect
on inner planets cannot be modelled by a uniform ring. The resulting
list contained 379 asteroids. Then, 100 asteroids whose masses were
determined negative (though always within uncertainty) were excluded
from the model. Such negative masses are an artifact of the
least-squares method and they come from the fact that there is not
enough observational data to properly determine them.

None of the 279 masses were fixed in planetary solutions in this work;
rather, all of them were determined along with other planetary
parameters by a weighted least-squares method enhanced with Tikhonov
regularization \citep{Kuchynka2013,KanPavlov}.
The \textit{a priori} masses and uncertainties used in
the regularization were taken from different sources. The masses of 17
asteroids are known with small uncertainties, thanks to observations
of spacecraft or their natural satellites (comparing to EPM2017,
binary asteroid Euphrosyne was added to the list). For 77 asteroids,
the \textit{a priori} values were the ones determined from observed
deflections of orbits of other asteroids during close approach. For
the remaining 285 asteroids, the estimates were taken based on
diameters observed in infrared \citep{IRAS,Masiero2011} and densities
assigned to the three taxonomic classes.

\subsection{Solutions}
Three solutions were made, with identical sets of observations and parameters,
differing only in solar wind delay model for the time span of 2006--2017.

\begin{itemize}
\item In Solution I, the delays were calculated in accordance with the
  simplest model of electron density, symmetric and stationary: $\Ne =
  C N_0 / r^2$, where $N_0$ (the reference electron density at $r=1$
  AU) was made equal to the 2006--2017 time-average of OMNI data, which
  was found to be 5.97 cm$^{-3}$. The formal error of this average value
  was calculated to be 0.07 cm$^{-3}$ from the formal errors of the proton
  density and alpha/proton ratio: $\sigma(N_\mathrm{p})$ and $\sigma(N_{\mathrm{a}}/N_{\mathrm{p}})$.
  The single electron density factor $C$ (which works as
  the factor of delay, see the $\tau_\mathrm{cor}$ term
  in \Cref{eq:tau_integral}) was fit to observations.
  Ideally, $C$ would be equal to 1; the purpose of $C$ is to account
  for uncertainty in the observed electron density, and also
  for the fact that the assumption of stationary electron
  density in 2006--2017 is not very realistic.
\item In Solution II, the delays were calculated in accordance with a
  symmetric, but non-stationary electron density model: $\Ne = C
  N_1(t) / r^2$, where the function $N_1(t)$ is is equal to the
  smoothed electron density values derived from OMNI (see \Cref{sec:insitu-data} and \Cref{fig:omni-savgol}).
  The single factor $C$ was fit to observations.
\item In Solution III, the delays were calculated in accordance with the
  WSA-ENLIL model of electron density, scaled so that the
  (365-day rolling average of)
  electron density near Earth is equal to the smoothed OMNI values.
  Similarly to Solutions I and II, a single factor $C$ of the electron
  density was fit to observations.
\end{itemize}

The values of the fitted electron density factor $C$ in the three
solutions are listed in \Cref{tbl:results-wrms}. While its reference
value is 1 in all solutions, there is a difference in how it is fitted.
In Solution I, the $C$ was
not constrained, since the assumption of stationary electron
density in 2006--2017 is already not very realistic. In Solutions II
and III, the Tikhonov regularization was used to effectively constrain
$C$ so that its \textit{a priori} standard deviation is 1.17\%.
The value  was taken equal to the relative standard deviation of
the electron density in the OMNI data (see the description of the
Solution I above).

Pre-2006 spacecraft ranges, which include MGS and parts of Odyssey and
MEX, were processed with the symmetric stationary solar wind
model, with different factors fit for different periods of time
between Sun--Mars conjunctions. One factor was fit for
2002/01/01---2003/12/31, another for 2004/01/01---2006/09/29.

A different factor was fit for pre-2002 MGS ranges. The same factor
was used for pre-1990 radar observations of Mars, Mercury, and Venus,
although they are hardly sensitive to delay in solar wind because of
low accuracy.

\section{Results}
\label{sec:results}
\begin{figure*}
  \centering
  \includegraphics{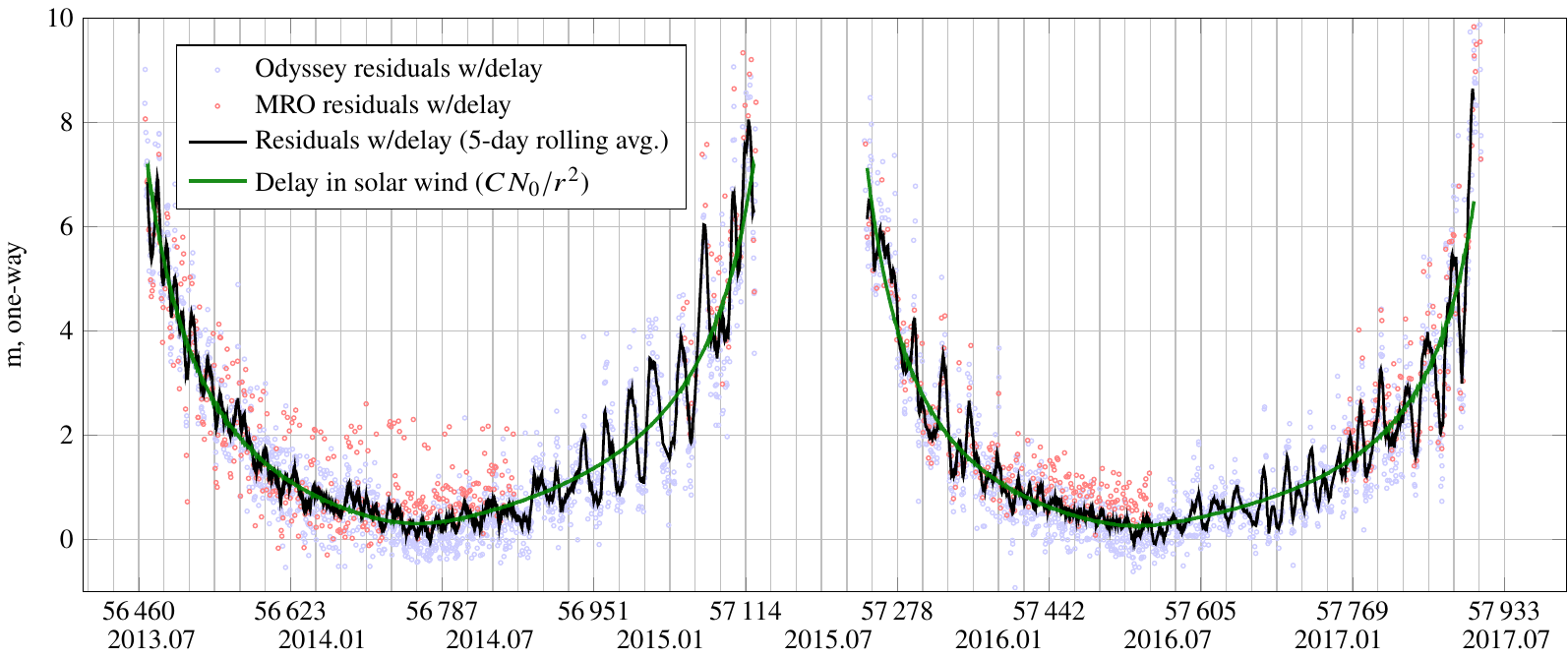}

  \includegraphics{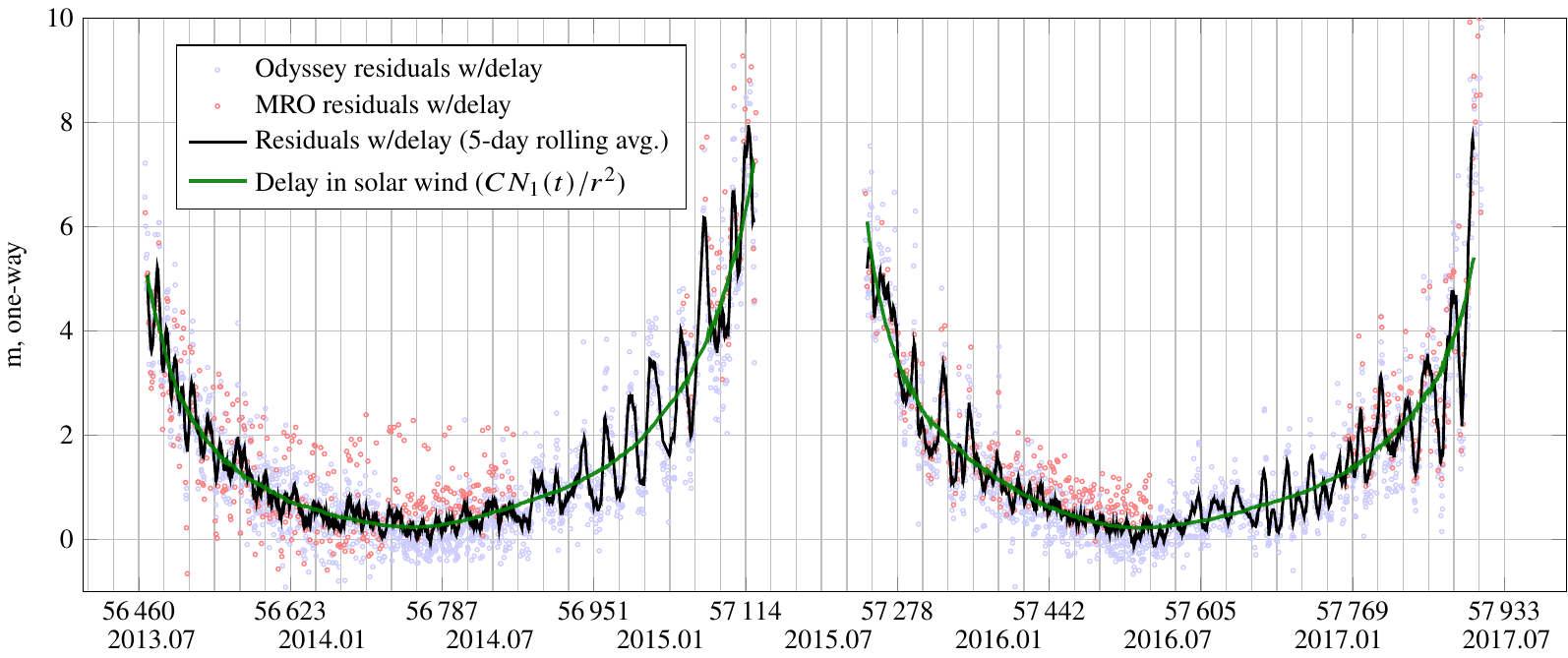}

  \includegraphics{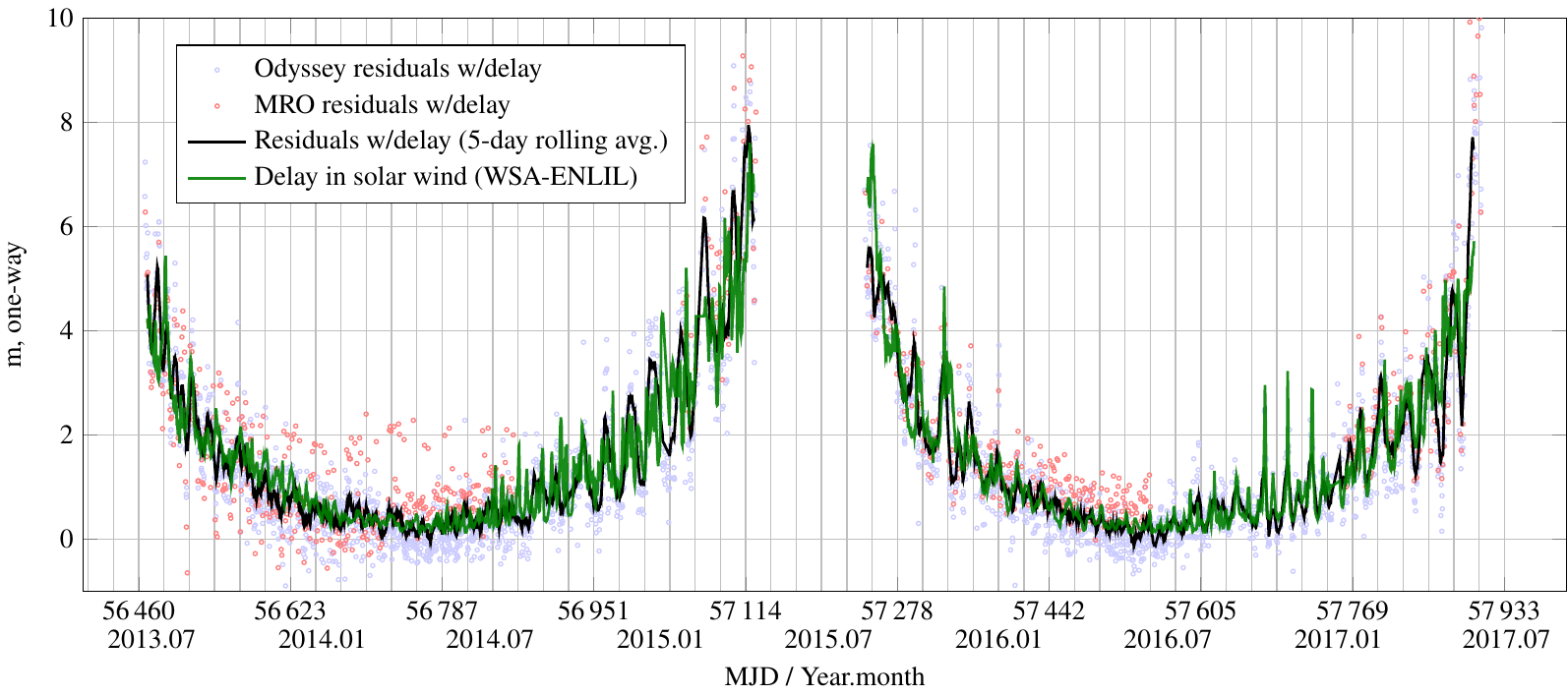}

  \caption{
    Odyssey and MRO range residuals without correction for delay in solar wind
    (blue and red points, respectively),
    their 5-day rolling averages (black curve),
    and model values (green curve) of delays in solar wind fitted
    to observations via the factor $C$. The top plot corresponds to Solution I
    ($C N_0/r^2$ model), the middle plot corresponds to Solution II ($C N_1(t)/r^2$ model),
    and the bottom plot corresponds to Solution III (WSA-ENLIL model).
    The time span is from mid-2013 to
    mid-2017.  The vertical grid lines are placed 27.2753 days apart so as to
    assist in examining the periodic variations due to the Sun's rotation.  }
  \label{fig:oc-mro-ody-1r2}
\end{figure*}

\Cref{fig:oc-mro-ody-1r2} shows the residual (postfit) differences of
Odyssey (blue points) and MRO (red points) ranges plus the values of
the model delay in solar wind. For better visibility, only
ranges from mid-2013 to mid-2017 are shown. The model delay in solar wind
is shown separately with green curves, so that one can see how the model
delay curve is fitted to observations.

The black curves show the time-averaged spacecraft range residuals.
They have visible medium-term
variations with period close to 27 days (the period changes as Earth
and Mars move on their orbits; the vertical lines on \Cref{fig:oc-mro-ody-1r2}
are separated by the solar rotation period). The amplitude of
variations increases near Sun--Mars conjunctions, which is expected
because during the conjunctions, Earth--Mars signals pass close to the Sun,
and at small distances from the Sun the electron density is higher than
at large distances. Observations with
signal passing closer than $30R_{\odot}$ to the Sun are not shown.

On the bottom plot of \Cref{fig:oc-mro-ody-1r2} it can be seen that
the WSA-ENLIL solar wind delay model expectedly has 27-day variations,
too. In some cases there is a good match between peaks in the model
and in the residuals, but often it goes ``out of sync'' or the
amplitudes become very different.
One explanation for this is the unfortunate fact that all the GONG
magnetograms are obtained from Earth---and the daily-updated map of
the magnetic field that is fed to the ENLIL's MHD equations is correct
only on the visible side of the Sun. On the nonvisible side, it is a
prediction, often not a very accurate one.

Before 2017, we can also attribute the differences between WSA-ENLIL
model data and real spacecraft delays to the GONGb magnetograms
containing non-solar magnetic field bias (see
\Cref{sec:enlil_data}). For 2017, when GONGz magnetograms were used,
we support the first explanation with anecdotal evidence. A good match
between WSA-ENLIL model and spacecraft data can be clearly seen on the
bottom plot of \Cref{fig:oc-mro-ody-1r2} in the first three months of
2017, while in the following three months, this is no longer the case.
\Cref{fig:enlil_sample_map} shows two example maps for the two
quarters of 2017. One can see that on the first map, the corotating
structures that intersect the Earth--Mars line originate from the
visible half of the Sun's surface; while on the second map, the
Earth--Mars line has changed in the way that it is also intersected by
a corotating structure whose origin is on the nonvisible half. (If the
pictures were animated, the structures would rotate anticlockwise in
time.) We admit that this can not be accepted as strong evidence and
hope for a future possibility to analyze post-2017 spacecraft
data together with WSA-ENLIL simulations based on properly fixed GONGz
magnetograms.

\begin{figure*}
  \centering
  \includegraphics{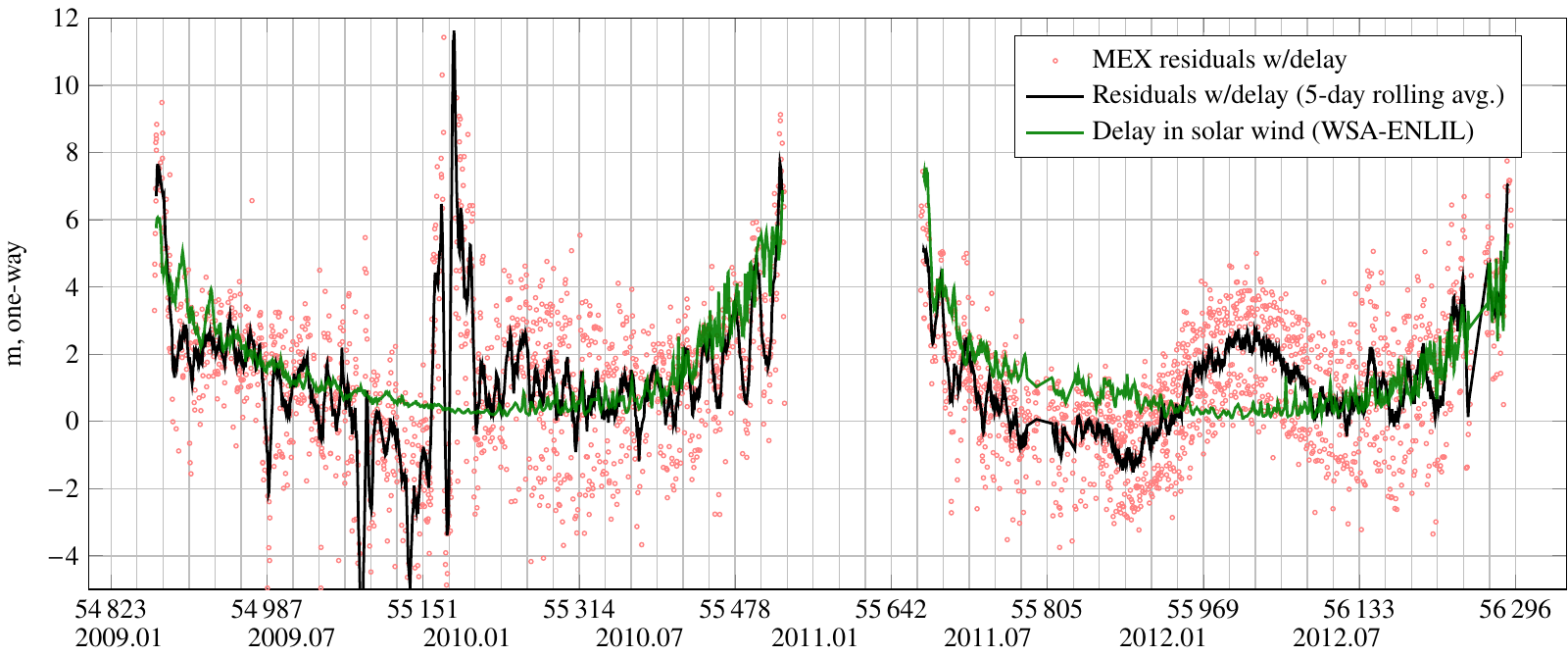}

  \caption{
    MEX range residuals without correction for delay in solar wind (red points)
    their 5-day rolling averages (black curve),
    and model values (green curve) of solar wind delays fitted
    to observations via the factor $C$. The plot corresponds to Solution III (WSA-ENLIL model).
    The time span is from mid-2013 to
    mid-2017. The vertical grid lines are placed 27.2753 days apart so as to
    assist in examining the periodic variations due to the Sun's rotation.  }
  \label{fig:oc-mex}
\end{figure*}

The Mars Express data are shown in \Cref{fig:oc-mex}. One can see,
apart from the expected peaks with a period of approximately 27
days, long-term variations that far exceed the peaks. The cause of the
variations certainly cannot be an inaccuracy of the EPM's dynamical
model of motion of Mars, or reductions, or solar wind delay model,
because there are no variations of that magnitude in the Odyssey and MRO
residuals.  It is also unlikely that the variations come from a radio
observatory on Earth: same Deep Space Network stations observe Mars
Express, Odyssey and MRO spacecraft. The most reasonable explanation
would be that there are problems in orbit determination for the spacecraft.
In any case, the comparison of solar wind delay models against the MEX
observations currently makes little scientific sense, at least with
the ranges that pass farther than 30 $R_{\odot}$ from the Sun.

The statistics of the postfit results and values of important
parameters are given in \Cref{tbl:results-wrms}. Solution II
(time-dependent symmetric model) is slightly better than Solution I
(stationary symmetric model) overall by the reduced $\chi^2$ metric
and particularly better near conjunctions.
Solution III (WSA-ENLIL-based solar wind delay model) is no better than
Solution II, most probably because of the aforementioned problems with
magnetograms.

\begin{table}
    \centering
    \caption{Weighted root-mean-square (WRMS) statistics and values of
      important parameters in the three solutions. The listed uncertainties are $3\sigma$.}
    \label{tbl:results}
    \begin{tabular}{lrrr}
        \hline
        &
        \textbf{Solution I} &
        \textbf{Solution II} &
        \textbf{Solution III} \\
        & $1/r^2$ & $1/r^2$ + OMNI & WSA-ENLIL \\
        \hline
        \textbf{WRMS}         &         &         &         \\
        \hspace{1em} Odyssey  & 0.564 m & 0.563 m & 0.563 m \\
        \hspace{1em} --- conj.$^\dagger$ & 1.16 m & 1.15 m & 1.15 m \\
        \hspace{1em} MRO & 0.619 m & 0.620 m & 0.619 m \\
        \hspace{1em}  & 1.20 m & 1.19 m & 1.19 m \\
        \hspace{1em} MEX & 2.46 m & 2.46 m & 2.46 m \\
        \hspace{1em}  & 1.96 m & 1.96 m & 1.95 m \\
        \hspace{1em} VEX & 6.24 m & 6.04 m & 6.04 m \\
        Density    & $1.31 \pm 0.81$ & $1.037 \pm 0.03$ & $1.037 \pm 0.03$ \\[-0.5ex]
        factor $C$ &        &        &        \\
        $\Delta GM_{\odot}^\ddagger,\ \mathrm{km}^3/\mathrm{s}^2$ & $0.972 \pm 0.526$ 
                                    & $1.457 \pm 0.463$
                                    & $1.470 \pm 0.463$ \\
        Variance of & 1.1127 & 1.1124 & 1.1124 \\[-0.5ex]
        unit weight      &        &        &        \\[-0.5ex]
        (reduced $\chi^2$)      &        &        &        \\ \hline \hline
        \multicolumn{4}{p{8cm}}{
          $^\dagger$ ``conj.'' (conjunction) observations are those where
          the closest distance from the Earth--Mars signal to the Sun
          is between 30 and 60 solar radii
        }\\
        \multicolumn{4}{l}{
          $^\ddagger$ $\Delta GM_{\odot} =
                      GM_{\odot} - 132712440041\ \mathrm{km}^3/\mathrm{s}^2$
        }
    \end{tabular}
    \label{tbl:results-wrms}
\end{table}

In Solutions II and III, the electron density delay factor $C$ was
constrained by Tikhonov regularization to be close to the reference
value of 1. An additional experiment (not shown) was made where the
Tikhonov regularization was not applied to $C$ in Solution II.  The
resulting value of $C$ turned out to be $1.35 \pm 0.81$ (as opposed to
$1.037 \pm 0.03$ in ordinary Solution II). This confirms that
proper usage of absolute \textit{in situ} measurements of
electron density really helps to bring the solution, including the
$GM_{\odot}$ estimate, closer to reality.  Also, the constraint on the
factor helps to reduce the uncertainty of $GM_{\odot}$ by $\sim 12$\%
(the uncertainty is 0.526 in Solution I and 0.463 in Solution II).

Another important improvement of Solution II over previous EPM
solutions is that only a single parameter related to solar wind is fit in
the solution, as compared to more than 50 in EPM2017
(see $B$ and $\dot B$ in \cref{eq:bbdot}, per-year, per-planet)
or 10 ($B$ per-conjunction from 2002 to 2018 and single $B$ prior to 2002)
in a later EPM version that was not released \citep{Pavlov2019,Pitjeva2021}.

\section{Conclusion and future work}
Conclusion:
\begin{itemize}
  \item Interplanetary spacecraft ranging observations are sensitive
    to signal delay due to solar wind and, in particular, to its medium-term
    variations caused by the rotation of the Sun, and to its long-term
    variations caused by changes in Sun's activity.
  \item The ENLIL model of solar wind, built on daily magnetograms
    done by the GONG network, can serve as a basis for a numerical
    solar wind delay model. Such a model, though, can have or not
    have a good match to the MRO and Odyssey ranges. Spacecraft ranges
    that pass through the parts of the electron density map that are
    effectively built from visible disk magnetograms have
    (with otherwise equal conditions) better residuals than
    the ranges that pass through the parts of the map that
    are built from older magnetograms rotated to the present date.
  \item Mathematical model of ephemeris can benefit from taking
    into account a solar wind delay model that has
    long-term variability determined from the OMNI dataset of
    \textit{in situ} observations.
  \item Constraining the (usually unconstrained) electron density
    factor in a planetary ephemeris solution to a reasonable range
    allows to reduce the error in the determined gravitational
    parameter of the Sun by $\sim$ 12\%.
  \item Using the WSA-ENLIL-based solar wind delay model to account
    for medium-term variations can improve the residuals in individual
    cases, but does not improve the solution overall.
\end{itemize}

Interplanetary spacecraft ranging observations are sensitive enough to
electron density so that they are used in this work to perform
comparative analysis and validation of solar wind models.
They can be used for validation of future solar wind models;
and potentially they can be used as a source of data for solar wind models.

\section*{Acknowledgements}
The authors would like to thank their colleagues from the IAA RAS:
Elena Pitjeva for continuous support of this work and a lot of
helpful advice regarding the planetary ephemeris solutions, and
Margarita Kan for implementing the Tikhonov regularization
method in the ERA-8 software and putting together the list of
\textit{a priori} masses of asteroids.

Planetary observations, including spacecraft ranges that are essential for this
work, were collected at the NASA SSD webpage supported by William Folkner
and at the Geoazur website supported by Agn{\`e}s Fienga.

The authors acknowledge use of NASA/GSFC's Space Physics Data
Facility's OMNIWeb service, and OMNI data, and are thankful to the
OMNIWeb team.

Solar wind simulation results have been provided by
the Community Coordinated Modeling Center at Goddard Space Flight Center
through their public ``Runs on Request'' system.

The numerical heliospheric code ENLIL has been developed by Dusan
Odstrcil who is currently at the George Mason University.

The WSA model, which gives the boundary conditions for the ENLIL
model, was developed by Yi-Ming Wang and Neil R. Sheeley Jr. at the
Naval Research Laboratory and later improved by Nick Arge, who worked at
the National Oceanic and Atmospheric Administration in Boulder at the
time and now works at the Goddard Space Flight Center.

The authors are indebted to Dusan Odstrcil for dedicated numerical
experiments and detailed explanations of inner workings of the ENLIL
model, to Leila Mays for organizing the numerical simulations at the
CCMC, and to Maria Kuznetsova for leading the CCMC itself.

\section*{Data Availability}
\label{sec:data_availability}

The Mars Global Surveyor (MGS), Mars Reconnaissance Orbiter (MRO),
and Odyssey spacecraft ranging data underlying this article
are available at the webpage of Solar System Dynamics (SSD) group at NASA
JPL\footnote{\url{https://ssd.jpl.nasa.gov/planets/obs_data.html}}.

The extended set of ranges for MRO and Odyssey up to
the end of 2017 was kindly provided by William Folkner of the SSD group
by permission and may be shared on reasonable request to the SSD team
of NASA JPL.

Mars Express (MEX) and Venus Express (VEX) ranges were downloaded from the Geoazur
website\footnote{\url{http://www.geoazur.fr/astrogeo/?href=observations/base}}.

The WSA-ENLIL data were produced by request via the ``Runs on Request'' service
at the Community Coordinated Modeling Center (CCMC) website%
\footnote{\url{http://ccmc.gsfc.nasa.gov}}.
The data are available for download through the ``Runs on Request'' service%
\footnote{\url{http://ccmc.gsfc.nasa.gov/ungrouped/SH/Solar_main.php}},
the names of the runs are:
\begin{itemize}
  \item \href{http://ccmc.gsfc.nasa.gov/database_SH/Dan_Aksim_122319_SH_2.php}{Dan\_Aksim\_122319\_SH\_2} (2006)
  \item \href{http://ccmc.gsfc.nasa.gov/database_SH/Dan_Aksim_122319_SH_3.php}{Dan\_Aksim\_122319\_SH\_3} (2007)
  \item \href{http://ccmc.gsfc.nasa.gov/database_SH/Dan_Aksim_122319_SH_4.php}{Dan\_Aksim\_122319\_SH\_4} (2008)
  \item \href{http://ccmc.gsfc.nasa.gov/database_SH/Dan_Aksim_122319_SH_5.php}{Dan\_Aksim\_122319\_SH\_5} (2009)
  \item \href{http://ccmc.gsfc.nasa.gov/database_SH/Dan_Aksim_122319_SH_6.php}{Dan\_Aksim\_122319\_SH\_6} (2010)
  \item \href{http://ccmc.gsfc.nasa.gov/database_SH/Dan_Aksim_122319_SH_7.php}{Dan\_Aksim\_122319\_SH\_7} (2011)
  \item \href{http://ccmc.gsfc.nasa.gov/database_SH/Dan_Aksim_122319_SH_8.php}{Dan\_Aksim\_122319\_SH\_8} (2012)
  \item \href{http://ccmc.gsfc.nasa.gov/database_SH/Dan_Aksim_122319_SH_9.php}{Dan\_Aksim\_122319\_SH\_9} (2013)
  \item \href{http://ccmc.gsfc.nasa.gov/database_SH/Dan_Aksim_120920_SH_1.php}{Dan\_Aksim\_120920\_SH\_1} (2014)
  \item \href{http://ccmc.gsfc.nasa.gov/database_SH/Dan_Aksim_122319_SH_11.php}{Dan\_Aksim\_122319\_SH\_11} (2015)
  \item \href{http://ccmc.gsfc.nasa.gov/database_SH/Dan_Aksim_122319_SH_12.php}{Dan\_Aksim\_122319\_SH\_12} (2016)
  \item \href{http://ccmc.gsfc.nasa.gov/database_SH/Dan_Aksim_112119_SH_3.php}{Dan\_Aksim\_112119\_SH\_3} (2017)
\end{itemize}

The OMNI \insitu data are available publicly through the OMNIWeb website%
\footnote{\url{http://omniweb.gsfc.nasa.gov}}.
Specifically, the file ``omni\_01\_av.dat'', containing daily averages
of solar wind parameters, comes from the low-resolution version%
\footnote{\url{http://spdf.gsfc.nasa.gov/pub/data/omni/low_res_omni/}}
of the OMNI dataset.

\bibliographystyle{mnras}
\bibliography{references}

\appendix

\section{Radio ranging residuals and determination of parameters}

\label{sec:residuals}
Radio signal travels twice the distance between the observatory on Earth
and the spacecraft. The time delay is being measured. The ranging
residual is the difference between the ``observed'' delay (normal
point, see Sec.~\ref{sec:spacecraft-ranges}) and the model
(``computed'') value of the delay.
The process of calculation of ``observed minus computed'' value
of the delay is called reduction. 

The precision of the normal point is sub-meter; for the model value to
match that precision, its formula must account for: the speed of
light, relativistic signal delay caused by Sun and other massive
bodies, proper time of observer, delay in troposphere and ionosphere,
delay in solar wind. All the mentioned delays are to be
calculated twice --- for the up and down legs of the signal.
A~detailed description can be found e.g. in \cite{Moyer2003}.
For each normal point, two equations are solved iteratively:
one to find the time of retransmission of signal by the spacecraft;
and the other to find the time of reception of the retransmitted
signal.

Observations must also be corrected for the spacecraft transponder
delay (the short moment of time between reception of signal by the
spacecraft and transmission of the return signal). There is always
some delay; it is measured before launch, but varies due to space
conditions as the spacecraft makes its way to the planet. The
transponder delay thus itself must be determined from radio ranges.

In addition to transponder delays, there are biases that come from
miscalibrations on radio observatories. Following the decision from
\citet{Kuchynka2012}, two sets of biases were determined for Deep
Space Network (DSN) stations: one for MGS and Odyssey spacecraft,
another for the MRO spacecraft.

In short, the following system of equations has to be solved for
each observable:

\newcommand{\PBCRS}{{\bm p}_\mathrm{BCRS}}
\newcommand{\SBCRS}{{\bm s}_\mathrm{BCRS}}
\newcommand{\Dgrav}{\Delta_\mathrm{grav}}
\newcommand{\Datm}{\Delta_\mathrm{atm}}
\newcommand{\Dsol}{\Delta_\mathrm{sol}}

\begin{equation}\label{traveltime}
\begin{cases}
t_2 - t_1 = \frac{|\PBCRS(t_2) - \SBCRS(t_1)|}{c} +
              \Dgrav(t_1, t_2) + \Datm(t_1, t_2) + \Dsol(t_1, t_2), \\
t_3 - t_2 = \frac{|\PBCRS(t_3) - \SBCRS(t_2)|}{c} +
              \Dgrav(t_3, t_2) + \Datm(t_3, t_2)  + \Dsol(t_3, t_2),
\end{cases}
\end{equation}

\noindent where $t_1$, $t_2$, and $t_3$ are the times of emission,
reflection, and reception of the signal in the Barycentric Dynamical
Time (TDB), a timescale that represents time at the barycenter
of the Solar system. Usually, a normal point contains $t_1$ in UTC, and
conversion to TDB is required.  $\PBCRS$ and $\SBCRS$ are the
positions of the center of the planet and the reference point of the
radio observatory (station) in the barycentric celestial reference
system (BCRS).

The ``computed'' value of the delay is
\begin{equation}\label{computed}
  \begin{aligned}
    C =  c \left[t_3 - t_1\vphantom{x^x}\right. + &\ \textrm{[TT}-\textrm{TDB]}(t_3, \SBCRS(t_3))\ -\\
      & \left.\vphantom{x^x}\textrm{[TT}-\textrm{TDB]}(t_1, \SBCRS(t_1))\right] + b,
  \end{aligned}
\end{equation}

\noindent where $b$ is the sum of transponder delay and the bias
(converted to units of distance). TT stands for \textit{terrestrial time}---a
timescale based on atomic clock and representing time on Earth's
surface. For calculating $[\textrm{TT}-\textrm{TDB}]$ at time $t$ and
point $\SBCRS(t)$, a theoretical equation is used, which can be found
for instance in \citep[eq. 2-23]{Moyer2003}. The geocentric terms of
the equation are integrated along with the Solar system equations and
stored in ephemeris; just one topocentric term is taken into account
in \Cref{computed}: $\left(\bm{\dot r}_E(t)\cdot(\SBCRS(t) -
\bm{r}_\mathrm{E}(t))\right)/c^2$.

One has to transform the position of the station in the terrestrial
reference frame to geocentric celestial reference frame (GCRS)
at times $t_1$ and $t_3$. The standard IAU2000/2006
precession-nutation model \citep{WallaceCapitaine} and
IERS EOP series \citep{bizouard17} are used for this.
Also, displacements of the station due to solid Earth tides and pole
tides are to be calculated according to the IERS Conventions 2010
\citep{iers2010}.

$\Dgrav$ is the delay of the signal due to spacetime curvature; it was
calculated in this work with a theoretical approximation
\citep{kopeikin}; delays from the Sun, Earth, the Moon, Jupiter, and
Saturn were added up.

$\Datm$ is the delay of the signal in ionosphere and troposphere,
usually calculated according to respective IERS models~\citep{iers2010}.
In this work, the given observations (normal points) were already corrected
for the atmospheric delay.

$\Dsol$ is the delay of the signal in solar wind, the subject of
this work.

\medskip

While the ``observed'' values are known and do not change, the
``computed'' values depend on model parameters (see
\Cref{sec:dynmodel}). The model parameters are to be fitted so that
the differences between the observed and computed, i.e. residuals,
are minimized. The values of residuals obtained after the fitting
(minimization) are called \textit{postfit} residuals.

The fitting is done via the Gauss--Newton method. The following function
is subject to minimization over $\bm\beta$:

\begin{equation}\label{eq:minimized}
  S({\bm\beta}) = \sum_{i=1}^m \frac{r_i({\bm\beta})^2}{\sigma_i^2} +
  \sum_{j=1}^n \frac{(\beta_j - \mu_j)^2}{\varepsilon_j^2},
\end{equation}

\noindent where $\bm\beta$ is the vector of all model parameters,
$r_i$ is the $i$-th residual ($m$ being the number of observations),
and $\sigma_i$ is the \textit{a priori} error (formal error)
that represents the standard deviation of the noise of the $i$-th observation.

In reality, the ``true'' thermal noise of measurement is hard to
separate from other systematic errors unaccounted for in the model.
The main source of those errors is the solar wind plasma itself;
also there can be errors from incompleteness of the set of perturbing
asteroids (see sec.~\ref{sec:asteroids}) or from wrong estimates of
asteroid masses or orbits. Finally, systematic errors (other than
the aforementioned constant offsets) can
come from hardware on the spacecraft or at the radio observatory.
A frequent approach to overcome this problem, which is also used in this
work, is to choose the formal errors to be somewhat smaller than
\textit{a posteriori} estimates of the standard deviation of
noise. Those estimates come from postfit residuals (see
Sec.~\ref{sec:results}). The decision to make the formal errors
smaller than the postfit residuals was made in order to account for
systematic errors in processing otherwise unaccounted for.

The second term of \cref{eq:minimized} comes from the Tikhonov
regularization modification of least-squares. $\mu_j$ is the \textit{a
  priori} value of the parameter $\beta_j$, while $\varepsilon_j$ is the
prior uncertainty of that value (taking $\varepsilon_j = \infty$ will mean
effectively no prior value).

\section{Difficulties of determination of~solar wind model parameters
         simultaneously with other parameters of~planetary ephemeris solution}
\label{sec:difficulties}
\begin{figure}
  \includegraphics{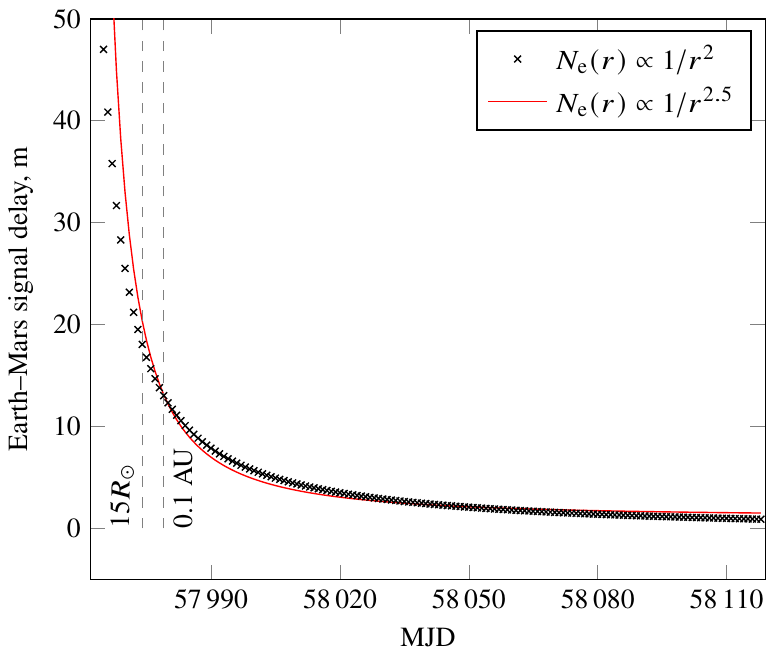}
  \caption{
    Earth--Mars signal delays with $1/r^2$ and $1/r^{2.5}$ density models.
    The multiplier (0.37) and the additive constant (1.16 m) for the $1/r^{2.5}$ curve are chosen
    so that the curve fits the $1/r^2$ curve on 57974--58119 interval, which corresponds
    to Earth--Mars signal passing farther than $15R_\odot$ from the Sun.
  }
  \label{fig:1r2_vs_1r25}
\end{figure}

Apart from solar wind plasma delay, observations of radio signal delays
always contain unknowns that are not directly related to solar wind,
but have a similar effect on the ranges. It is important not
to be overly confident in e.g. the determined value of the power
$\epsilon$ in the $1/r^\epsilon$ model of solar wind density.
\Cref{fig:1r2_vs_1r25} provides a synthetic example with
$\epsilon=2.5$.  Let us assume that the ``real'' solar wind behaves
according to the $1/r^2$ model, but we fit the $1/r^{2.5}$ model to
observations. That is, we must fit the electron density factor, but also an
offset (bias) that is inherent to spacecraft observations. It turns
out that the $1/r^{2.5}$ model can be made virtually equivalent to
the $1/r^2$ one as long as a factor and a bias are fitted, and
signal paths that pass closer than 15 solar radii to the Sun are
excluded from the analysis due to their bad quality, as they normally
are in ephemeris solutions (in fact, the barrier is often 30 solar
radii or more).

\begin{figure}
  \includegraphics{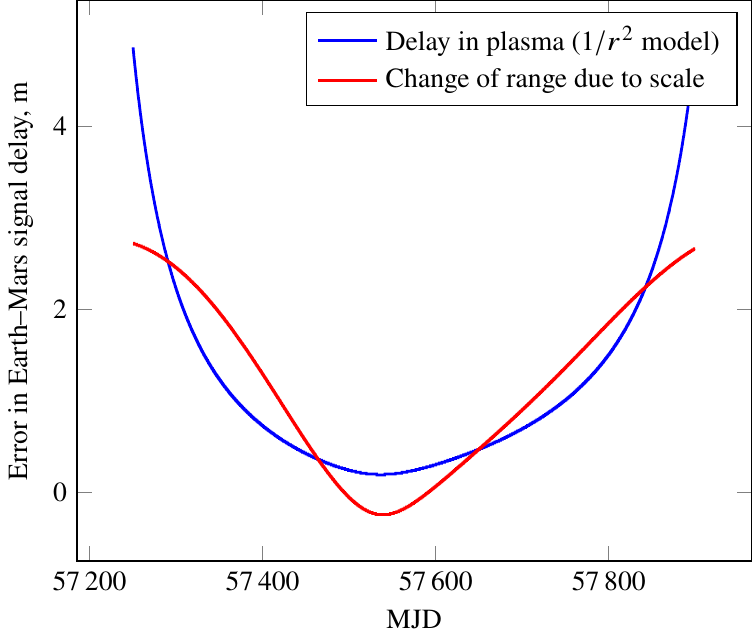}

  \caption{Red curve (change of Earth--Mars range due to scale) fitted to the blue curve (model values of delay in solar wind), with a factor of ($1 + 9.67\times 10^{-12}$) and an offset of 0.97 m. The two curves are correlated at 87\%.}
  \label{fig:corona_vs_au}
\end{figure}

Another example concerns the fitted electron density factor
itself. Even if the power of $r$ in the model is correct, it is
important not to be overly confident in the determined factor. In a
planetary ephemeris solution, the $GM_{\odot}$ is an unknown
parameter that has to be determined from observations. In EPM, the
$GM_{\odot}$ is not routinely determined directly; a distance scale
parameter is determined instead. In any case, an incorrect value of
the $GM_{\odot}$ adds an error to the computed range, proportional
to the range itself. The delay of radio signal in solar wind is
largest near Sun--Mars conjunctions; it so happens that the
Earth--Mars distance follows the same pattern. That, together with the
unknown bias that also has to be fitted, makes the two sources of
error not perfectly separable, as illustrated by a synthetic example
on \Cref{fig:corona_vs_au}.
Let us assume that the blue curve is the ``real'' delay in solar wind.
If no reduction for solar wind delay was made in the residuals,
the residuals will lie on the blue curve. But if, instead of taking into
account the solar wind delay, we decide to fit the distance
scale factor and bias, we will end up subtracting the red curve
from the residuals. The blue curve is not very far off from the blue curve.
This leads to the conclusion that an incorrectly determined value of the
electron density factor can be partially absorbed by a small scale
factor (and hence a small change in the mass of the Sun) and small offset.

\bsp  %
\label{lastpage}
\end{document}